\documentclass[fleqn,usenatbib]{mnras}

\usepackage{newtxtext,newtxmath}
\usepackage[T1]{fontenc}
\usepackage{color, soul}

\DeclareRobustCommand{\VAN}[3]{#2}
\let\VANthebibliography\thebibliography
\def\thebibliography{\DeclareRobustCommand{\VAN}[3]{##3}\VANthebibliography}

\usepackage{graphicx}	% Including figure files
\usepackage{amsmath}	% Advanced maths commands
\usepackage[dvipsnames]{xcolor}
\usepackage{comment}
\usepackage[flushleft]{threeparttable}
\usepackage{ulem}

\defcitealias{wvm+22}{Paper~\uppercase\expandafter{\romannumeral1}}
\title[Dual-frequency scattering study of PSR~J0826+2637]{Pulsar Scintillation Studies with LOFAR: \uppercase\expandafter{\romannumeral2}. Dual-frequency scattering study of PSR~J0826+2637 with LOFAR and NenuFAR}

\author[Z. Wu et al.]{
Ziwei Wu $^{1,2}$, \thanks{E-mail: wuzw@bao.ac.cn}
William A.\ Coles $^{3}$,
Joris P.\ W.\ Verbiest $^{2, 4}$,
Krishnakumar Moochickal Ambalappat $^{4,2}$, 
\newauthor
Caterina Tiburzi $^{5}$,
Jean-Mathias Grie{\ss}meier $^{6,7}$,
Robert A.\ Main $^{4}$,
Yulan Liu  $^{1,2}$,
Michael Kramer $^{4,8}$,
\newauthor
Olaf Wucknitz $^{4}$,
Nataliya Porayko $^{4}$,
Stefan Os\l owski $^{9}$,
Ann-Sofie Bak Nielsen $^{4,2}$,
Julian Y.\ Donner $^{4,2}$,
\newauthor
Matthias Hoeft $^{10}$,
Marcus Brüggen $^{11}$,
Christian Vocks $^{12}$,
Ralf-Jürgen Dettmar $^{13}$,
Gilles Theureau $^{6,7,14}$,
\newauthor
Maciej Serylak $^{15,16}$,
Vladislav Kondratiev $^{17}$,
James W.\ McKee $^{18, 19}$,
Golam M.\ Shaifullah $^{20, 21, 5}$,
\newauthor
Ihor P.\ Kravtsov $^{22,23}$,
Vyacheslav V.\ Zakharenko $^{22}$,
Oleg Ulyanov $^{22}$,
Olexandr O.\ Konovalenko $^{22}$
\newauthor
Philippe Zarka $^{23, 24}$,
Baptiste Cecconi $^{23, 24}$,
Léon V.\ E.\ Koopmans $^{25}$,
and Stéphane Corbel $^{24, 26}$ \\
$^{1}$ National Astronomical Observatories, Chinese Academy of Sciences, 20A Datun Road, Chaoyang District, Beijing 100101, China \\
$^{2}$ Fakult\"at f\"ur Physik, Universit\"at Bielefeld, Postfach 100131, 33501 Bielefeld, Germany\\
$^{3}$ Electrical and Computer Engineering, University of California, San Diego, 92093, USA \\
$^{4}$ Max-Planck-Institut f\"ur Radioastronomie, Auf dem H\"ugel 69, 53121 Bonn, Germany \\
$^{5}$ INAF, Osservatorio Astronomico di Cagliari, Via della Scienza 5, 09047 Selargius, Italy \\
$^{6}$ LPC2E - Universit\'{e} d'Orl\'{e}ans / CNRS, France\\
$^{7}$ Nan\c{c}ay Radioastronomy Observatory (ORN), Observatoire de Paris - CNRS/INSU, USR 704 - Univ. Orl\'{e}ans, OSUC, Route de Souesmes, 18330 Nan\c{c}ay, France\\
$^{8}$ Jodrell Bank Centre for Astrophysics, The University of Manchester, Oxford M13 9PL, United Kingdom\\
$^{9}$ Manly Astrophysics, 15/41-42 East Esplanade, Manly, NSW 2095, Australia\\
$^{10}$ Thüringer Landessternwarte, Sternwarte 5, 07778 Tautenburg, Germany \\
$^{11}$ Hamburger Sternwarte, University of Hamburg, Gojenbergsweg 112, 21029 Hamburg, Germany\\
$^{12}$ Leibniz-Institut f\"ur Astrophysik Potsdam (AIP), An der Sternwarte 16, 14482 Potsdam, Germany\\
$^{13}$ Ruhr-Universität Bochum, Fakultät für Physik und Astronomie, Astronomisches Institut, 44780 Bochum, Germany \\
$^{14}$ Laboratoire Univers et Th{\'e}ories, Observatoire de Paris, Universit\'e PSL, Universit{\'e} Paris Cit{\'e}, CNRS, F-92190 Meudon, France\\
$^{15}$ SKA Observatory, Jodrell Bank, Lower Withington, Macclesfield, SK11 9FT, United Kingdom\\
$^{16}$ Department of Physics and Astronomy, University of the Western Cape, Bellville, Cape Town 7535, South Africa\\
$^{17}$ ASTRON, the Netherlands Institute for Radio Astronomy, Postbus 2, 7990 AA Dwingeloo, The Netherlands\\
$^{18}$ E.A. Milne Centre for Astrophysics, University of Hull, Cottingham Road, Kingston-upon-Hull, HU6 7RX, UK \\
$^{19}$ Centre of Excellence for Data Science, Artificial Intelligence and Modelling (DAIM), University of Hull, Cottingham Road, Kingston-upon-Hull, HU6 7RX, UK \\
$^{20}$ Dipartimento di Fisica ``G. Occhialini'', Universit\`a di Milano-Bicocca, Piazza della Scienza 3, 20126 Milano, Italy \\
$^{21}$ INFN, Sezione di Milano-Bicocca, Piazza della Scienza 3, I-20126 Milano, Italy \\
$^{22}$ Institute of Radio Astronomy of NAS of Ukraine, 4 Mystetstv St, 61002, Kharkiv, Ukraine\\
$^{23}$ LESIA, Observatoire de Paris, Université PSL, Sorbonne Université, Université Paris Cité, CNRS, 92190 Meudon, France \\
$^{24}$ USN, Observatoire de Paris, Université PSL, Univ Orléans, CNRS, 18330 Nançay, France \\
$^{25}$ Kapteyn Astronomical Institute, University of Groningen, P.O.Box 800, 9700AV Groningen, The Netherlands \\
$^{26}$ Université Paris Cité, Université Paris-Saclay, CEA, CNRS, AIM, F-91191 Gif-sur-Yvette, France
}

\date{Accepted XXX. Received YYY; in original form ZZZ}

\pubyear{2015}

\begin{document}
\label{firstpage}
\pagerange{\pageref{firstpage}--\pageref{lastpage}}
\maketitle

\begin{abstract}
Interstellar scattering (ISS) of radio pulsar emission can be used as a probe of the ionised interstellar medium (IISM) and causes corruptions in pulsar timing experiments. Two types of ISS phenomena (intensity scintillation and pulse broadening) are caused by electron density fluctuations on small scales (< 0.01 AU). 
Theory predicts that these are related, and both have been widely employed to study the properties of the IISM. Larger scales ($\sim$1-100\,AU) cause measurable changes in dispersion and these can be correlated with ISS observations to estimate the fluctuation spectrum over a very wide scale range.
IISM measurements can often be modeled by a homogeneous power-law spatial spectrum of electron density with the Kolmogorov ($-11/3$) spectral exponent.
Here we aim to test the validity of using the Kolmogorov exponent with PSR~J0826+2637. We do so using observations of intensity scintillation, pulse broadening and dispersion variations across a wide fractional bandwidth (20 -- 180\,MHz).
We present that the frequency dependence of the intensity scintillation in the high frequency band matches the expectations of a Kolmogorov spectral exponent but the pulse broadening in the low frequency band does not change as rapidly as predicted with this assumption. We show that this behavior is due to an inhomogeneity in the scattering region, specifically that the scattering is dominated by a region of transverse size $\sim$40\,AU. The power spectrum of the electron density, however, maintains the Kolmogorov spectral exponent from spatial scales of 5$\times10^{-6}$\,AU to $\sim$100\,AU. 
\end{abstract}


%%%%%%%%%%%%%%%%%%%%%%%%%%%%%%%%%%%%%%%%%%%%%%%%%%

%%%%%%%%%%%%%%%%% BODY OF PAPER %%%%%%%%%%%%%%%%%%

\section{Introduction}
Pulsars are rapidly rotating neutron stars that emit compact beams of radio waves from their magnetic poles. Their rotation rate is extremely stable and can be measured with high precision. This permits a number of applications including an effort to detect and characterize ultra-low-frequency gravitational waves by observing their effects on pulsar timing arrays \citep[PTAs, see, e.g.,][and references therein]{vob21}.
However, propagation effects imparted by the ionized interstellar medium (IISM) on pulsar signals are a source of timing noise that would substantially worsen the sensitivity of PTAs if not corrected \citep[see, e.g.,][and references therein]{vs18}.

Plasma density irregularities in the IISM have been modeled as a homogeneous three-dimensional spatial power spectrum in the form \citep{r77, r90}
\begin{equation}
P_{\rm n_e}(q) = C_{\rm n_e}^{2}\left(q^{2}+l_{\rm outer}^{-2}\right)^{-\beta/2}\exp\left(-q^{2}l_{\rm inner}^{2}/2\right),
\end{equation}
where $C_{\rm n_e}^{2}$ defines the level of the density spectrum of the IISM, 
$q$ is the three-dimensional wave number, 
$\beta$ is the spectral exponent, 
and $l_{\rm inner}$ 
and $l_{\rm outer}$ are the inner and outer scales of the density fluctuations, respectively. In the range 
($l_{\rm inner} \ll 1/q \ll l_{\rm outer}$) which is analogous to the inertial range of neutral turbulence,
 the form of the three-dimensional power spectrum can be simplified to $P_{\rm n_e} \approx C_{\rm n_e}^{2} q^{-\beta}$. The well-known Kolmogorov spectral exponent is $\beta = 11/3$. The Kolmogorov exponent was derived from a dimensional analysis for neutral turbulence and there is very little theoretical support for this exponent in an astrophysical plasma. There is general observational support \citep{ars95}, but there are also observations which are inconsistent with the homogeneous Kolmogorov model \citep{gkk+17}. Here we will show that some apparently inconsistent observations are consistent with the Kolmogorov exponent but not with a homogeneous scattering medium. 

We will present near-simultaneous observations of PSR~J0826+2637, a nearby (500\,pc) low dispersion measure (DM, 19.5 pc/cm$^3$) young pulsar, from the LOw-Frequency ARray (LOFAR) High Band Antennae (HBA) and the New extension in Nan\c{c}ay upgrading LOFAR (NenuFAR) which are used for scattering studies of the small-scale structure in a wide frequency range (20--180\,MHz). 
We also present a DM time series  using HBA  observations centered on $\sim$150\,MHz, which is used to study the larger-scale structure.
This work has been organized in the following manner: in Section~2 we outline the necessary scintillation theory; in Section~3 we describe our observations and data processing; 
in Section~4 we show the analysis and results. 
Section~5 contains our conclusions. 

\section{Scattering Theory}

The physics of scattering of pulsar radiation by turbulent interstellar plasmas has been studied by several authors \citep[see, e.g.,][and references therein]{r77, r90}. 
The primary scattering mechanism of pulsar signals is diffraction caused by random fluctuations in the refractive index of the IISM. This causes the pulsar radiation to arrive at the observer as an angular spectrum of plane waves, in which higher angles correspond to a longer delay. The pulse is thus broadened and develops a quasi-exponential tail. To first order the angular spectrum can be approximated as gaussian and, if the scattering is localized in a thin region, the pulse is broadened with a half-exponential shape. This approximation is adequate to describe the half power width of the angular spectrum and the pulse, but in fact both have power-law tails and these are very important in creating "scintillation arcs".

The statistic best suited for analyzing the ISS phenomena and the related DM variations is the phase structure function $D_{\rm \phi} (s) = \langle (\phi(r) - \phi(r+s))^2 \rangle$ where $\phi(r)$ is the phase on the geometrical path from the source to the observer at transverse position $r$. In a power-law medium with $2<\beta<4$ the structure function is also power-law: $D_{\rm \phi}(s) \sim s^{\beta-2}$.

The autocorrelation of the electric field is $\rho_{\rm e(s)} = \exp(-0.5 D_{\rm{\phi}}(s))$. Its Fourier transform is the angular scattering distribution $B(\theta)$. When the intensity scintillations are strong their autocorrelation is $\rho_{\rm{i}}(s) = |\rho_{\rm{e}}(s)|^2$. Thus the 1/e scale of the ISS is the scale $s_{\rm 0}$ at which $D(s) = 1$. The width of $B(\theta)$ is $\theta_{\rm 0} = 1/k s_{\rm 0}$ where $k = 2\pi/\lambda$ is the wavenumber. If $\beta = 4$ both $\rho_{\rm i}\, {\rm and}\,B(\theta)$ are gaussian. When $\beta = 11/3$ they are nearly gaussian and we often approximate their widths using a gaussian model.

If the scattering comes from a compact region on the line of sight, which is often referred to as a ``thin screen'', then the pulse broadening with scattering angle is given by $\tau = 0.5 \theta^2 Z_{\rm e}/c$ where $Z_{\rm e}$ is the effective distance of the screen. In the gaussian approximation, the pulse broadening function (PBF) is then:
$\rm{PBF(t)} = \exp\left(-t/ \tau_{\rm sc}\right)$ \citep{r77,rom86}.

Interference between components of the angular spectrum causes intensity scintillations. When these are observed in a dynamic spectrum they have a characteristic width in both time $\tau_{\rm 0}$ and frequency $\Delta\nu_{\rm d}$. The autocorrelation in time is just the spatial correlation converted by the velocity of the line of sight through the scattering medium $\rho_{\rm i}(s = V_{\rm eff} t)$, so $s_{\rm o} = V_{\rm eff}\tau_{\rm 0}$.
The autocorrelation in frequency is the Fourier transform of the PBF so that its width $\Delta\nu_{\rm d}$ is inversely related to $\tau_{\rm sc}$. In the gaussian approximation, which we will use here \citep{cr98}:
\begin{equation}
\label{eq:corrleation}
2\pi\,\tau_{\rm sc}\,\Delta\nu_{\rm d} = 1.
\end{equation}

The dependence of scintillation on the wavelength $\lambda$ in a homogeneous medium with a power-law spectrum is also power-law if the intensity scintillations are very strong, i.e.\ if their bandwidth $\Delta\nu_{\rm d} \ll 1$. The electric field coherence scale
$s_{\rm 0} \sim \lambda^{-2/(\beta-2)}$, the rms scattering angle $\theta_{\rm 0} \sim \lambda^{2/(\beta-2)+1}$ and the scattering time $\tau_{\rm sc} \sim \theta_0^2 \sim \lambda^{4/(\beta-2)+2}$. If $\beta = 11/3$ then
$\tau_{\rm sc} \sim \lambda^{4.4}$ and $\Delta\nu_{\rm d} \sim \lambda^{-4.4}$ If $\beta = 4$ as would occur with $s_{\rm inner} > s_{\rm 0}$ then the exponent ($\alpha$) changes to 4.0, whereas for a flatter $\beta = 3$ (such as occurs in the solar wind near the Sun), the exponent $\alpha = 6$. In a homogeneous medium $\alpha < 4$ is not possible.

Measurement of $\tau_{\rm sc}$ or $\Delta\nu_{\rm d}$ at different wavelengths are the source of most (but not all) inconsistencies  with the homogeneous Kolmogorov model 
\citep[e.g.,][]{bts+19,kmj+19,lvm+22}. This is the case with our observations of PSR~J0826+2637. When the scattering medium is not homogeneous one must distinguish between the angular scattering distribution and the angular spectrum of radiation seen by the observer. In the simple case of a scattering ``cloud'' of radius $R_{\rm cloud}$ the observed angular spectrum is limited to $\theta_{\rm obs} < R_{\rm cloud}/D_{\rm eff}$. As the wavelength increases $\theta_{\rm 0}$ will continue to increase as $\lambda^{2/(\beta-2)+1}$ but $\theta_{\rm obs}$ will eventually reach the limit of $R_{\rm cloud}/D_{\rm eff}$.
This will limit both $\tau_{\rm sc}$ and $\Delta\nu_{\rm d}$ causing the exponent $\alpha$ to be less than 4. 
We note that several scattering screen models including circular screen with finite radius have been discussed in \cite{cl01}.

\section{Observations and data processing} 
Typical pulse widths are $\sim$\,ms, so it is difficult to measure $\tau_{\rm sc} < 1$\,ms. The $\Delta \nu_{\rm d}$ corresponding to $\tau_{\rm sc} > 1$\,ms is so small it is difficult to measure. 
So it is unlikely that one can measure both $\tau_{\rm sc}\, {\rm and}\, \Delta \nu_{\rm d}$ at the same observing frequency.
In this work intensity scintillations from the LOFAR HBA are used to estimate $\Delta \nu_{\rm d}$ and pulse-profile evolution studies from NenuFAR\footnote{\url{https://nenufar.obs-nancay.fr/en/astronomer}} \citep[][Zarka et al.\ in prep.]{bgt+21, Zdt+20} are used to estimate $\tau_{\rm sc}$.
The Modified Julian Date (MJD), the length, time and frequency resolution of the selected archives are summarized in Table~\ref{tab:data}.

\begin{table*}
\begin{center}
\caption{Summary of observations and scattering parameters of PSR~J0826+2637.}
\begin{tabular}{lcccccccc}
\hline \hline
Telescope & MJD range & $N_{\rm{obs}}$ & Length & $\Delta f$ & $\Delta t$ & Parameter & $\alpha$  & Frequency\\
        &     &                &(mins) & (kHz)      &  (s)       &    &       &  (MHz)\\
\hline 
FR606   & 59101 & 1 & 60   & 0.6  & 10 & $\Delta \nu_{\rm{d}}$ & 4.58(16) & 120 - 180\\
NenuFAR & 59087 & 1 & 21.5 & 195 & 10 & $\tau_{\rm{sc}}$ & 2.7(1)   & 10 - 85  \\
GLOW    & 56500-59213 & 470&$\sim$60 & 195 & 10 & DM & -  & 120 - 180 \\
\hline
\end{tabular}
\label{tab:data}
\end{center}
\begin{tablenotes}
\small
\item {\bf Notes:} Given are the telescope name; the time span of the observations; the number of observations $\rm{N_{obs}}$; the observation length; the frequency resolution $\Delta f$; the time resolution $\Delta t$; 
the determined IISM parameters ($\Delta \nu_{\rm{d}}$, $\tau_{\rm sc}$ and DM are the scintillation bandwidth, the pulse broadening time, and the dispersion measure, respectively), the derived power-law index $\alpha$ and the frequency range of data. Values in brackets are the uncertainty in the last digit quoted; these uncertainties correspond to the formal 1-$\sigma$ error bar.
\end{tablenotes}
\end{table*}

\subsection{Intensity Scintillation Analysis}
\begin{figure}
        \centering
        \includegraphics[width=0.99\linewidth]{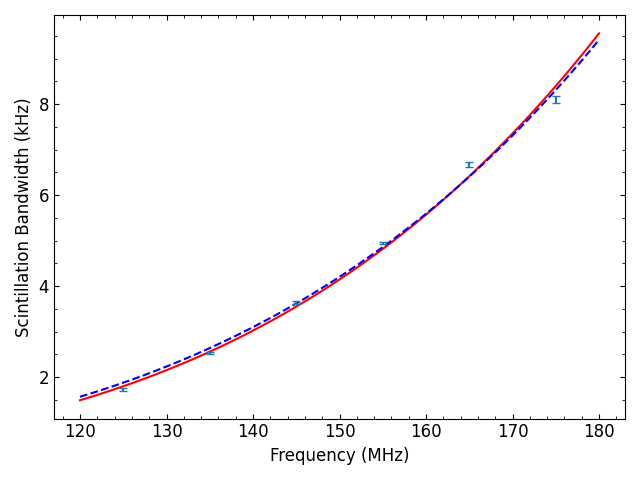}
        \caption{Measured scintillation bandwidth $\Delta\nu_{\rm{d}}$ vs frequency for LOFAR HBA observations. The fitted (red) line has an exponent 4.6. The blue dashed line has the Kolmogorov exponent $\alpha = 4.4$.}
        \label{fig:LOFAR_BW_freq}
    \end{figure}

The LOFAR HBA data recording and initial processing were identical to those described in \citet{dvt+19}: raw data from the telescope were processed by the \textsc{dspsr} package \citep{vb11} and written out in \textsc{psrfits} format \citep{hvm04}. Subsequent processing was carried out with the \textsc{psrchive} package \citep{vdo12}.
Radio frequency interference (RFI) is excised with the \textsc{iterative\_cleaner}\footnote{Available from \url{https://github.com/larskuenkel/iterative_cleaner}} in this work, which is a modification of the \textsc{surgical} method included in the RFI cleaner of the \textsc{coastguard} pulsar-data analysis package \citep{lkg+16}.
The frequency and time resolution of data from the International LOFAR Station in Nan\c{c}ay (FR606, \citealt{bgt+20}) were set to 0.6\,kHz and 10 seconds, respectively, in order to resolve the scintles.
The measured $\Delta\nu_{\rm d}$ is plotted vs frequency in
Figure~\ref{fig:LOFAR_BW_freq}. The best-fit exponent, shown by the solid red line, is $\alpha$ = 4.58$\pm$0.16, but the Kolmogorov exponent (4.4) is consistent with the error bars as shown by the dashed blue line.

\subsection{DM Analysis}

Our DM analysis is based on HBA data from five German stations of the International LOFAR Telescope \citep[ILT,][]{vwg+13}, taken between 05 March 2013 and 21 September 2021.
For the monitoring observations used in this work, the five LOFAR stations of the German LOng Wavelength (GLOW) consortium, located in Effelsberg (telescope identifier DE601, 75 observations), Tautenburg (DE603, 288 observations), Potsdam-Bornim (DE604, 12 observations), Jülich (DE605, 13 observations) and Norderstedt (DE609, 82 observations), are used as individual stand-alone telescopes, not connected to the ILT network, as described in detail by
\cite{dvt+19}.
In the end, the time and frequency resolution of GLOW data are $\sim$10\,s 
and $\sim$0.195\,MHz, respectively, with 1024 pulse phase bins. 
The details of our DM determination are identical to the description by \cite{tsb+21}. 

\begin{figure} 
\centering
\includegraphics[width=0.99\linewidth]{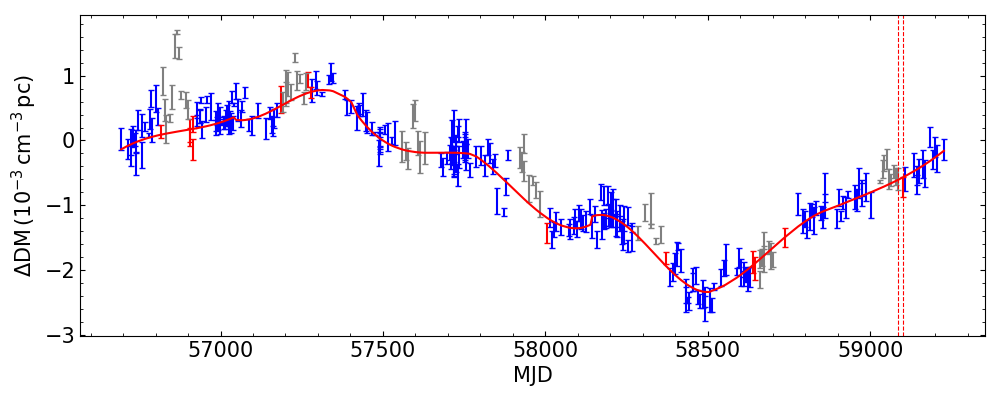}
\caption{$\rm{DM}$ time series in the direction of PSR~J0826+2637 measured by LOFAR. A DM baseline of 19.47921 $\rm{cm^{-3}\,pc}$ is subtracted. Observations closer than 40$^\circ$ from the Sun are gray error bars. Observations outside of 50$^\circ$ are blue error bars and those between 40$^\circ$ and 50$^\circ$ are red error bars. The ISM contribution estimated using the techniques in \citealt{tsb+21} is shown in red. Two dashed lines indicate the epochs MJD~59087 and MJD~59101, respectively.}
\label{fig:DM_obs}
\end{figure}

The ecliptic latitude of PSR~J0826+2637 is only 7.24$^\circ$ so the solar wind contribution to DM is significant for several months of the year. We use the techniques of \cite{tsb+21} to estimate the ISM component of DM, then we remove all observations with a solar elongation $<50^\circ$ and we remove outliers with respect to the ISM component greater than 3 sigma. The results are shown as blue error bars in Figure~\ref{fig:DM_obs}.
The times of the observations used for our scattering study are shown as vertical dashed lines. They are taken at solar elongation of $\sim 40^\circ$ at which a solar wind contribution is possible. We augmented the plot with red error bars for observations between 40$^\circ$ and 50$^\circ$
and it is clear that there are no significant DM variations, as measured at 150 MHz, around these times. 
However the observations occur during a significant positive DM gradient which will shift the lines of sight at LOFAR and NenuFAR differentially. 

The DM is a column density so it is directly proportional to the additional phase due to the plasma on the line of sight. The constant of proportionality $C_{dm2\phi} = 0.830\pi10^{10}/\nu_{\rm MHz} \rm{rad}\, \rm{pc}\, \rm{cm}^{3}$.

The DM gradient, $\sim$5.8333 $\times 10^{-5}$ pc cm$^{-3}$/AU, can be converted to a phase gradient and the corresponding angular shift $\theta_{\rm g} = \nabla\phi / k$ calculated. We find that $\theta_{\rm g} = 0.28\, \theta_{\rm 0}$ at the highest NenuFAR frequency of 46.6\,MHz. This shift is in the direction of the velocity so the observations of $\tau_{\rm sc}$ are in a region of somewhat higher DM. However the scattering disc for the LOFAR HBA observations remains within the scattering disc for NenuFAR observations.
Note that $\theta_{\rm g} \propto \lambda^2$ whereas $\theta_o \propto \lambda^{2.2}$ so the relative importance of $\theta_{\rm g}$ decreases slowly with $\lambda$.

\subsection{Pulse Profile Analysis}
PSR~J0826+2637 is part of a long-term monitoring programme with NenuFAR.
The NenuFAR data used for pulse profile studies have 2048 pulse phase bins.
To correct dispersion in the narrow spectral channels, coherent dedispersion was applied in real time with the Low frequency Ultimate Pulsar Processing Instrument (LUPPI) described by \cite{bgt+21}. 
Due to the lack of a well-tested calibration scheme for NenuFAR data, we focus on the uncalibrated total-intensity data. 

To model the pulse profiles, the fitting model reported in \citet{kmn+15} is used. 
With the assumption of a simple thin screen model dominating the scattering \citep[e.g.,][]{wil72}, the observed pulse profile can be expressed as a convolution of the frequency-dependent intrinsic pulse shape $P_{i}(t, \nu)$ with the PBF(t), i.e. 
$\rm{P(t) = P_{i}(t,\nu) \ast PBF(t)}$, 
where $\ast$ denotes convolution. 
The pulse broadening by DM is negligible since coherent dedispersion has been applied.

\begin{figure}
\centering
\includegraphics[width=0.99\linewidth]{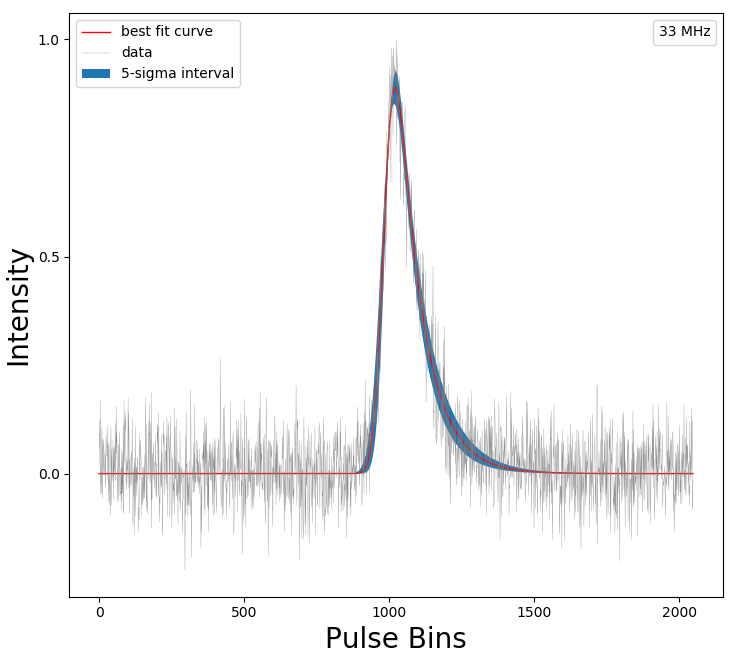}
\caption{Average total-intensity NenuFAR pulse profile (in grey) of PSR~J0826+2637 at 33$\pm$1\,MHz at epoch MJD~59087. The time and frequency resolution of NenuFAR data are $\sim$10 s and $\sim$0.195 MHz, respectively. The red curve represents the best-fitting model and the blue band is the uncertainty.}
\label{fig:fitting}
\end{figure}

PSR~J0826+2637 is known to have a three-component pulse profile consisting of main pulse, postcursor and an interpulse and exhibits a bright (B-mode) and a quiet emission mode (Q-mode) in LOFAR observations \citep{syh+15}.
In our NenuFAR observations, no nulling and Q-mode are detected and the postcursor and interpulse components are hidden in the noise (see Fig.~\ref{fig:fitting}). 
Thus, in this work, a single Gaussian profile model is used as the intrinsic pulse profile.

\begin{figure}
 \centering
 \includegraphics[width=0.99\linewidth]{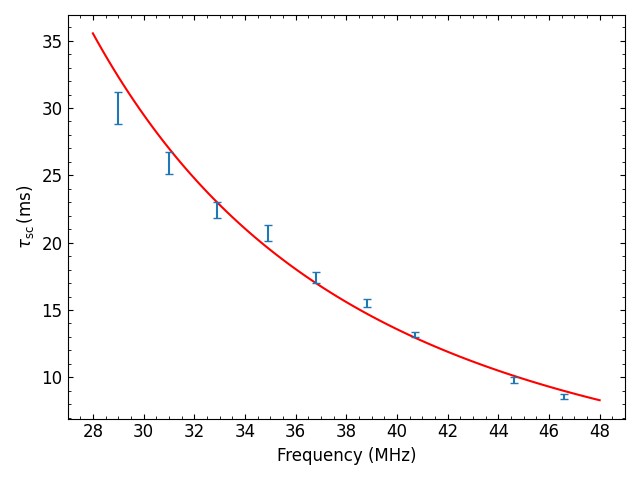} 
 \caption{The pulse broadening time $\rm{\tau_{sc}}$ versus observing frequency for PSR~J0826+2637. The red line has an exponent $\alpha = -2.7$.}
 \label{fig:Tsc_freq}
\end{figure}

An illustration to demonstrate the procedure used to estimate $\tau_{\rm{sc}}$ is given in Figure~\ref{fig:fitting}, which shows an example of PSR~J0826+2637 at $33\pm1\,$MHz.
Afterwards, we estimate $\alpha$ from $\tau_{\rm{sc}}$ derived from multiple sub-bands across our data (see Figure~\ref{fig:Tsc_freq}).
We note that some $\tau_{\rm{sc}}$ values are excluded from the subsequent analysis (specifically, for deriving the power-law index $\alpha$), in particular when $\tau_{\rm{sc}}$ is nearly equal to or smaller than the pulse width at the higher frequency bands. It is clear that the best power-law exponent is near 2.7$\pm$0.1, but that a power-law fit is marginal because neither extreme frequency is well matched. The scattering time observations are definitely inconsistent with the assumption of a thin screen and a homogeneous power-law scattering medium.

\subsection{Scintillation Arc}
Intensity scintillation can be observed to form a dynamic spectrum of intensity as a function of frequency and time $I(\nu,t)$. The 2D Fourier transform of this dynamic spectrum $S_{\rm sec}(\tau , f_{\rm{d}}) = F_2(I(\nu,t))$, is the ``secondary spectrum''. The dynamic spectrum is caused by interference between components of the scattered angular spectrum of plane waves. Those scattered plane waves each have a well-defined delay and Doppler shift with respect to an unscattered plane wave. The axes of $S_{\rm sec}$ are the differential delay $\tau_{\rm d}$ and differential Doppler shift $f_{\rm{d}}$, so $S_{\rm sec}$ is a distribution of the received angular spectrum in differential delay and differential Doppler.

Parabolic arcs can be observed in the $S_{\rm sec}$ only when the scattering is dominated by a compact region somewhere along the line of sight \citep{wms+04, crs+06}. In this case $\tau_{\rm d}$ and $f_{\rm d}$ are uniquely defined by the scattering angles of the two interfering waves $\boldsymbol{\theta_{\rm 1}}$ and $\boldsymbol{\theta_{\rm 2}}$ giving 
\begin{align}
    \tau_{\rm d} &= D\, ((1-s)/2cs)\,(\boldsymbol{\theta_{\rm 1}}^2 - \boldsymbol{\theta_{\rm 2}}^2)\,\,\,\rm{and}\\
    f_{\rm d} &= (1/\lambda s)\,(\boldsymbol{\theta_{\rm 1}} - \boldsymbol{\theta_{\rm 2}})\cdot \boldsymbol{{V}_{\rm{eff}}}\,\,\, \rm{where}\\
    \boldsymbol{V_{\rm{eff}}} &= (1-s) \boldsymbol{V_{\rm{p}}} + s \boldsymbol{V_{\rm{E}}} - \boldsymbol{{V}_{\rm{ISM}}}(s).
\end{align}
Here s is the fractional distance from the pulsar to the scattering screen. The pulsar velocity is 272 kms$^{-1}$ \citep{dgb+19} so, to first order, the contributions of $V_{\rm E}$ and $V_{\rm ISM}$ can be ignored.

In PSR~J0826+2637 we see a forward arc with its apex near the origin of $S_{\rm sec}$. Such arcs are caused by highly scattered plane waves interfering with waves closer to the origin of the angular spectrum. The ``boundary'' arc is defined by the maximum Doppler for a given delay, so $\boldsymbol{\theta_{\rm 2}} \rightarrow 0$ and $\boldsymbol{\theta_{\rm 1}}$ is parallel to $\boldsymbol{{V}_{\rm eff}}$. It is very useful to normalize $\tau_{\rm d}$ by the values they would have if $\theta_{\rm 1} = \theta_{\rm o}$, i.e. $\tau_{\rm sc}$ and $2 \pi s /\tau_{\rm o}$. This which gives
$\tau_{\rm dn} = (\theta_{\rm 1} / \theta_{\rm o})^2$ and $f_{\rm dn}$ = $\theta_{\rm 1}/\theta_{\rm o}$. The arc is then given by $\tau_{\rm dn} = f_{\rm dn}^2$ so the normalized arc curvature is unity.

When a phase gradient in the direction of the velocity is present in the scattering medium the entire angular spectrum is shifted by an angle $\theta_{\rm g}$ \citep{crs+06}. This does not alter the Doppler $f_{\rm dn}$ but it changes the normalized delay to
\begin{align}
    \tau_{\rm dn} &= f_{\rm dn}^2 + 2 (\theta_{\rm g}/\theta_{\rm o}) f_{\rm dn}\,\,\,{\rm and}\\
    \tau_{\rm d} &= \eta f_{\rm d}^2 + \gamma f_{\rm d}.
\end{align}
The parabolic arc still passes through the origin but its apex is shifted to $f_{\rm dn} = -\theta_{\rm g}/\theta_{\rm o}$. This factor increases from 0.28 at 46.6 MHz, to 0.354 at 150 MHz. %\jpwv{Is this described or derived anywhere (e.g.\ in any of the previously cited works by Cordes or Walker)?} \WC{Some of it is in Cordes et al, 2006 but $\gamma$ was not calculated explicitly} \jpwv{Thanks.}

\begin{figure}
\centering
\includegraphics[width=0.99\linewidth]{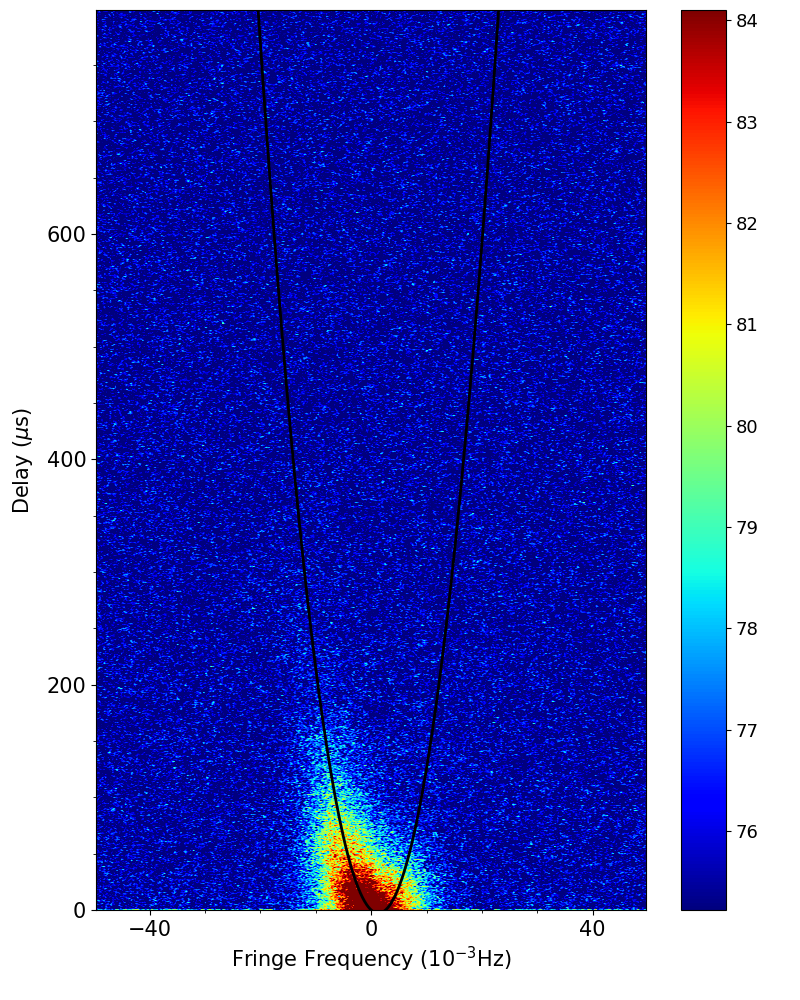}
\caption{Observed secondary spectrum of PSR~J0826+2637 with LOFAR in the frequency range 145 -- 155\,MHz at MJD 59101. The asymmetric parabola corrects for the phase gradient and implies s=0.56.}
\label{fig:new_arc}
\end{figure}

The secondary spectrum observed on MJD 59101 is shown in Figure~\ref{fig:new_arc} in observed units $\tau_{\rm d}$ and $f_{\rm d}$. The amplitude is $\log_{10}(S_{\rm sec})$. The arc is not sharply defined but $\eta = 1.7 s^3$ and $\gamma = -0.00441 s^2$ can be estimated. This provides an estimate of s = 0.56. If the phase gradient were not included one would estimate s = 0.66, a 20\% difference.
    
\section{Analysis and results}
To resolve the inconsistency between the scintillation bandwidth and scattering time observations, it is helpful to put both in  context with the larger scale $\Delta \rm{DM}(t)$ observations. This could be done with the power spectra, but it is more directly done using the structure function of DM. We first convert $\Delta \rm{DM}(t)$ to $\Delta \rm{DM}({r})$ using $\boldsymbol{r} = \boldsymbol{V_{\rm eff}}\,t$. The structure function of DM is $D_{\rm DM}(s) = \langle(\rm{DM}(r) - \rm{DM}(r+s))^2\rangle$. It can be directly compared with the $\tau_{\rm sc}\, {\rm and}\, \Delta\nu_{\rm d}$ measurements each of which provides a measurement of $s_{\rm 0}$. By definition, $D_\phi(s_0) = 1$. Since $\phi$ is directly related to DM this provides a measurement of $D_{\rm DM}(s_0)$ at a much smaller scale.
The phase $\phi$ equals $C_{dm2\phi}\, \rm{DM}$ where the constant $C_{dm2\phi} = 0.830\, \pi\, 10^{10}/\,\nu_{\rm MHz}$.

So $D_{\rm DM} = D_\phi\, C_{dm2\phi}^{-2}$.
In this way a sample of $D_\phi(s_0) = 1$ obtained from a bandwidth or a scattering time measurement, can be converted to $D_{\rm DM}(s_0) = C_{dm2\phi}^{-2}$. The value of $s_0$
can be obtained from $\theta_0 = (\tau_{\rm sc} 2 c / V_{\rm eff})^{0.5}$ and $s_0 = 1 / k \theta_0$.

\begin{figure}
        \centering
        \includegraphics[width=0.99\linewidth]{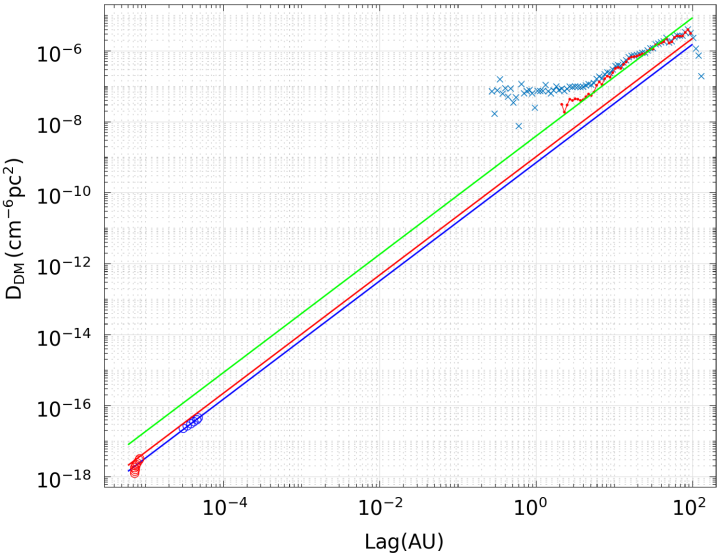}
        \caption{Structure function $D_{\rm DM}(s = V_{\rm eff} t)$. The blue x symbols are the raw estimates. The red filled circles have had the white noise contribution subtracted. The white noise subtraction is not reliable for $s < 2$ AU and $D_{\rm DM}$ is heavily biased by the limited data length for $s > 100$ AU, so the red line is limited to $2 < s < 100$ AU.
        The blue circles are from the LOFAR HBA measurements of $\Delta\nu_d$. The red squares are from the NenuFAR $\tau_{\rm sc}$ values. The straight red line is the best Kolmogorov model which fits the NenuFAR data. The straight blue line is the equivalent for the LOFAR HBA data. The straight green line best fits the observed $D_{\rm DM}$.}
        \label{fig:DDM_obs}
    \end{figure}

The structure function $D_{\rm DM}$ was obtained directly from the definition by calculating all the squared differences and binning them to obtain an average. We did not weight the squared differences by their white noise error bars because the white noise is not the primary source of estimation error. That is the finite number of independent estimates in that average. The estimate of $D_{\rm DM}$ directly from the $\Delta \rm{DM}(s = V_{\rm eff} t)$ time series is plotted in Figure~\ref{fig:DDM_obs}, with estimates from the HBA scintillation bandwidth measurements and the samples of $\tau_{\rm sc}$ from NenuFAR.

One can see that the Kolmogorov extrapolation of the large scales exceeds the NenuFAR 46.6 MHz estimate by a factor of 3.8. The LOFAR HBA estimates, which by themselves have the Kolmogorov exponent, fall below the large-scale extrapolation by a factor of 5.6. There are two points to be explained: the scattering observations come from a weaker scattering region than the average of the $\Delta \rm{DM}$ observations; and the NenuFAR observations cannot come from a thin homogeneous scattering screen. The scattering screen must be thin because an arc is observed, so it cannot be homogeneous. This inhomogeneity can also explain why the HBA observations come from a weaker scattering region.

A simple model of an inhomogeneity that would explain our observations, is that the scattering region has a finite extent of $\sim$40\,AU. At frequencies of $\ge 46.6$ MHz the scattering disc $\theta_0 D_{\rm eff}$ is smaller than 40\,AU so the scattering appears homogeneous. However at lower frequencies the scattering disc exceeds 40\,AU. 
This limits the angular spectrum received by the observer and thus $\tau_{\rm sc}$ fails to increase as expected. Inhomogeneity also explains why the scattering estimates of $D_{\rm DM}(s_0)$ are lower than the large scale average. Indeed the direct observations of $\Delta \rm{DM}(t)$ indicate that DM during the observations is lower than the average but increasing. The refractive gradient will cause the NenuFAR observations to be observed later than the LOFAR HBA observations and thus in a region of stronger scattering. The displacement of the scattering disc is only about 28\% of its width at 46.6 MHz, but the width of the scattering disc at 150 MHz is very much smaller. Its area is < 1\% of the area of the NenuFAR scattering disc, so the reduced scattering at 150 MHz could be due to a density variation on a scale of $\sim$\,2AU which would be invisible to the low frequencies.

\subsection{Finite Screen Model}

The angular spectrum scattered by the turbulent plasma is the same as the angular spectrum seen by the observer if the scattering medium is homogeneous. However if the scattering plasma does not fully occupy the scattering disc then the observed angular spectrum will be smaller than the scattered angular spectrum. This will increase $s_0\, {\rm and}\, \Delta\nu_{\rm d}$ but decrease $\tau_{\rm sc}$. As the size of the scattering disc increases at lower frequencies ($\propto \nu^{-2.2}$ for a Kolmogorov spectrum) the effect becomes more pronounced at lower frequencies. 

Clearly the behavior of, e.g. $\tau_{\rm sc}$, depends on the distribution of the plasma within the scattering disc. Here we choose the simple model of a gaussian plasma distribution $\rm{B(\theta)_{\rm{cloud}} = exp(-(\theta/\theta_{cloud})^2)}$ centered in a gaussian scattering disc $\rm{B(\theta)_{\rm{scat}} = exp(-(\theta/\theta_{scat})^{2})}$. The observer will then see  
\begin{equation}
B(\theta)_{\rm{obs}} = \rm{exp}(-[(\theta/\theta_{\rm{scat}})^{2} + (\theta/\theta_{\rm{cloud}})^2])
= \rm{exp}(-(\theta/\theta_{\rm{tot}})^2)
\end{equation}
where $\theta_{\rm{tot}}^{2} = (1/\theta_{\rm{scat}}^{2} + 1/\theta_{\rm{cloud}}^{2})^{-1}$.
In terms of measurable parameters, $\tau_{\rm sc}\, {\rm and}\, \Delta \nu_{\rm{d}}$, $\theta_0$ is replaced by $\theta_{\rm tot}$.

\begin{figure}
    \centering
    \includegraphics[width=0.99\linewidth]{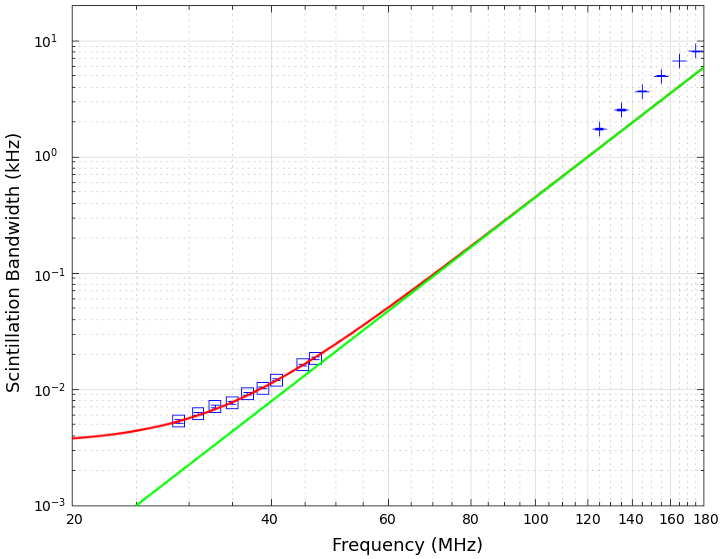}
    \caption{The scintillation bandwidth $\Delta \nu_{\rm{d}}$ as a function of observing frequency. The blue squares are calculated from the observed $\tau_{\rm sc}$ at NenuFAR. The solid red curve that passes through the NenuFAR observations is the finite screen model adjusted to best match them. It has a cloud size that matches the scattering disc at 33 MHz, i.e. $\sim\, 40\,$AU. The bandwidth computed for this model with a very large blob size is shown as a solid green line.The LOFAR HBA measurements of scintillation bandwidth are shown as blue + symbols.}
    \label{fig:band_data_model}
\end{figure}

The model is compared with the data in Figure~\ref{fig:band_data_model}. 
The data points marked with blue squares are the bandwidth calculated from the measured $\tau_{\rm sc}$ at NenuFar. The solid red curve that passes through the NenuFAR observations is the finite screen model adjusted to best match them. It has a cloud size that matches the scattering disc at 33 MHz, i.e. $\sim\, 40\,\rm{AU}$. The $D_{\rm DM}$ of this model is shown as a straight red line on Figure~\ref{fig:DDM_obs}. It lies a factor of 3.8 below that computed directly from the $\Delta DM$ observations in the range 2 to 100 AU. The bandwidth computed for this model with a very large blob size is shown as a solid green line. The LOFAR HBA bandwidth measurements are shown as blue + symbols. They lie a factor of 1.5 above the green line (i.e. lower in turbulence level).

\subsection{Summary}
The parabolic arc measurements show the scattering screen is relatively thin and located roughly midway between the pulsar and the Earth. The $\Delta \rm{DM}(t)$ observations show that the fluctuations, averaged over a 100 AU trajectory are Kolmogorov between scales of 2 AU and 100 AU. The structure function of DM, $D_{\rm{DM}}$, can be extrapolated down to scales of 5 $\times$ $10^{-6}$ AU  where it exceeds the estimated $D_{\rm{DM}}$ from the NenuFAR observations by a factor of 3.8 and that estimated from the LOFAR HBA observations by a factor of 5.7.

The LOFAR HBA estimates are consistent with a Kolmogorov exponent but at a lower level. The lower frequency NenuFAR observations cannot be explained by a thin scattering screen with homogeneous power-law spatial spectrum, but they can be modeled using a finite scattering screen with a spatial scale of $\sim$\,40\, AU. Thus all observations are consistent with a Kolmogorov scattering medium that is not stationary in variance on scales $\ge$40\,AU.

\subsection{Other Observations}
The frequency-dependence of $\tau_{\rm sc}$ and $\Delta\nu_{\rm d}$ for PSR~J0826+2637 has previously been studied: 
\cite{bts+19} found that the median value of $\alpha$ is 1.55$\pm$0.09 based on their long-term monitoring of $\tau_{\rm sc}$ with the Long Wavelength Array (LWA) at frequencies ranging between 44 and 75\,MHz; 
\cite{kmj+19} reported a value of 2.4$\pm$0.1 for $\alpha$ based on combined $\tau_{\rm sc}$ data from the LWA and the Ooty Radio Telescope in the frequency range of 32--62\,MHz; 
\citet{dlk13} found $\alpha$ = 3.94$\pm$0.36 from the measured $\Delta{\nu_{\rm{d}}}$ over a wide frequency band (300-1700\,MHz) at different epochs; and in \citetalias{wvm+22} we reported $\alpha$ = 4.12$\pm$0.16 from LOFAR HBA measurements of $\Delta{\nu_{\rm{d}}}$ at 150\,MHz.
Our analysis of $\tau_{\rm sc}$ gives $\alpha = 2.71\pm0.11$ (see Table~\ref{tab:data}).
In contrast, our analysis of LOFAR $\Delta\nu_{\rm d}$ taken on MJD 58820 results in $\alpha = 4.58\pm0.16$. It is clear that the low frequency observations are all affected by inhomogeneity. It is also clear that to make an accurate estimate of the spectral exponent one needs a longer spatial baseline. Comparison of scattering observations and DM observations provides a much more precise technique. 

\subsection{Other Sources of Error}
There are other possibilities summarized below that could bias estimates of the pulse broadening delay $\tau_{\rm{sc}}$: 
\begin{enumerate}
 \item The pulsar radio emission mechanism.
There are two components in the fitted model $P(t)$: $\tau_{\rm sc}$ and the intrinsic pulse profile.
The width of the intrinsic pulse profile is frequency-dependent, which is not fully understood yet \citep{lk12} and could affect the above measurements. 
\item Pulse profile with multiple components. 
The postcursor and interpulse components of PSR~J0826+2637 are not distinguished in our NenuFAR data due to the limited S/N. This could disturb the fitting procedure. 
In order to resolve postcursor and interpulse components of PSR~J0826+2637 with NenuFAR, the data with longer observation length are required. Moreover, other fitting procedures; e.g., the forward fitting method of \cite{gk16}, and the deconvolution method by \cite{bcc03} could provide more insight into pulsar profile analysis, particularly for pulsars with multi-component pulse profiles.
\end{enumerate}

\section{Conclusion}
We find that near-simultaneous observations of PSR~J0826+2637 from NenuFAR at 28 to 48 MHz and LOFAR from 120 to 180 MHz are not consistent with a homogeneous power-law scattering medium. The LOFAR measurements are internally consistent with a homogeneous Kolmogorov medium over a scattering disc of 2.5 AU in radius. The NenuFAR observations are consistent with a Kolmogorov scattering blob of $\sim$\,40\,AU in radius. However the scattering disc of the NenuFAR observations below 46 MHz is larger than 40 AU and this limits the scattering time at lower frequencies.

The non-standard frequency-dependence of the scattering properties presented here, could be a common feature particularly at very low observing frequencies, since similar abrupt inhomogeneities on AU scales have regularly been reported in observations of scintillation arcs \citep{srm+22}.
Further observations including other low frequency facilities, e.g. the Long Wavelength Array \citep{tek+12} the Murchison Widefield Array \citep{tgb+13} and the Ukrainian T-shaped Radio telescope \citep{zvk+13} could significantly improve the models of the physical size of scattering screens in the near future.

\section*{acknowledgements}
%personal
The authors thank George Hobbs for discussions and advice. 
We also thank the anonymous referee for the
valuable suggestions that improved this paper. 
ZW acknowledges support by Bielefelder Nachwuchsfonds through Abschlussstipendien für Promotionen.
This work is supported by National SKA Program of China No. 2020SKA0120200.
JPWV acknowledges support by the Deutsche Forschungsgemeinschaft (DFG) through the Heisenberg programme (Project No.\ 433075039).
YL acknowledges support from the China scholarship council (No. 201808510133).
J. W. McKee gratefully acknowledges support by the Natural Sciences and Engineering
Research Council of Canada (NSERC), [funding reference CITA 490888-16].
M.B. acknowledges support from the Deutsche Forschungsgemeinschaft under Germany's Excellence Strategy - EXC 2121 "Quantum Universe" -- 390833306.
%LOFAR
LOFAR \citep{vwg+13} is the Low Frequency Array designed and constructed by ASTRON. It has observing, data processing, and data storage facilities in several countries, that are owned by various parties (each with their own funding sources), and that are collectively operated by the ILT foundation under a joint scientific policy. The ILT resources have benefitted from the following recent major funding sources: CNRS-INSU, Observatoire de Paris and Universit\'{e} d'Orl\'{e}ans, France; BMBF, MIWF-NRW, MPG, Germany; Science Foundation Ireland (SFI), Department of Business, Enterprise and Innovation (DBEI), Ireland; NWO, The Netherlands; The Science and Technology Facilities Council, UK.
% GLOW-LOFAR introduction
This paper uses data obtained with the German LOFAR stations,
during station-owners time and ILT time allocated under project codes
LC0\_014, LC1\_048, LC2\_011, LC3\_029, LC4\_025, LT5\_001, LC9\_039, LT10\_014 and LT14\_006.
We made use of data from
% DE601
the Effelsberg (DE601) LOFAR station funded by the Max-Planck-Gesellschaft;
% DE602
the Unterweilenbach (DE602) LOFAR station funded by
the Max-Planck-Institut für Astrophysik, Garching;
% DE603
the Tautenburg (DE603) LOFAR station funded by the State of Thuringia,
supported by the European Union (EFRE) and the
Federal Ministry of Education and Research (BMBF) Verbundforschung
project D-LOFAR I (grant 05A08ST1);
% DE604
the Potsdam (DE604) LOFAR station funded by the
Leibniz-Institut für Astrophysik (AIP), Potsdam;
% DE605
the Jülich (DE605) LOFAR station supported by the
BMBF Verbundforschung project D-LOFAR I (grant 05A08LJ1);
% DE609
and the Norderstedt (DE609) LOFAR station funded by the
BMBF Verbundforschung project D-LOFAR II (grant 05A11LJ1).
% GLOW observations infrastructure
The observations of the German LOFAR stations
were carried out in stand-alone GLOW mode,
which is technically operated and supported by
the Max-Planck-Institut für Radioastronomie, the Forschungszentrum
Jülich and Bielefeld University. We acknowledge support and
operation of the GLOW network, computing and storage facilities by
the FZ-Jülich, the MPIfR and Bielefeld University and financial support
from BMBF D-LOFAR III (grant 05A14PBA) and D-LOFAR IV (grant 05A17PBA),
and by the states of Nordrhein-Westfalia and Hamburg.
We acknowledge the work of A.~Horneffer in setting up
the GLOW network and initial recording machines.
%FR606
LOFAR station FR606 is hosted by the Nan\c{c}ay Radio Observatory and
is operated by Paris Observatory, associated with the French Centre
National de la Recherche Scientifique (CNRS) and Universit\'{e} d'Orl\'{e}ans.
%NenuFAR
This paper is partially based on data obtained using 
the NenuFAR radio-telescope. The development of NenuFAR has
been supported by personnel and funding from: Station de 
Radioastronomie de Nançay, CNRS-INSU, Observatoire de Paris-PSL, Universit\'{e}
d’Orl\'{e}ans, Observatoire des Sciences de l’Univers en R\'{e}gion Centre,
R\'{e}gion Centre-Val de Loire, DIM-ACAV and DIM-ACAV+ of R\'{e}gion
Ile-de-France, Agence Nationale de la Recherche.
We acknowledge the use of the Nan\c{c}ay Data Center computing facility (CDN
- Centre de Donn\'{e}es de Nan\c{c}ay). The CDN is hosted by the Station de
Radioastronomie de Nançay in partnership with Observatoire de Paris,
Universit\'{e} d’Orl\'{e}ans, OSUC and the CNRS. The CDN is supported by the
R\'{e}gion Centre-Val de Loire, d\'{e}partement du Cher.
The Nan\c{c}ay Radio Observatory is operated by the Paris Observatory,
associated with the French Centre National de la Recherche Scientifique (CNRS).

\section*{Data Availability}
The data underlying this article will be shared on reasonable request
to the corresponding author.

\bibliography{journals,modrefs,psrrefs,crossrefs}
\bibliographystyle{mnras}

%\begin{thebibliography}{99}
%\end{thebibliography}

%%%%%%%%%%%%%%%%%%%%%%%%%%%%%%%%%%%%%%%%%%%%%%%%%%

%%%%%%%%%%%%%%%%% APPENDICES %%%%%%%%%%%%%%%%%%%%%

% Don't change these lines
\bsp	% typesetting comment
\label{lastpage}
\end{document}